\documentclass[conference]{IEEEtran}
\IEEEoverridecommandlockouts
\usepackage{cite}
\usepackage{amsmath,amssymb,amsfonts}
\usepackage{algorithmic}
\usepackage{graphicx}
\usepackage{textcomp}
\usepackage{xcolor}
\usepackage{verbatim}

\def\BibTeX{{\rm B\kern-.05em{\sc i\kern-.025em b}\kern-.08em
    T\kern-.1667em\lower.7ex\hbox{E}\kern-.125emX}}
\begin{document}

\title{Differential EEG Characteristics during Working Memory Encoding and Re-encoding\\
\thanks{This work was partly supported by the Institute for Information $\&$ Communications Technology Planning $\&$ Evaluation (IITP) grant funded by the Korea government (MSIT) (No. 2017-0-00451, Development of BCI based Brain and Cognitive Computing Technology for Recognizing User’s Intentions using Deep Learning; No. 2019-0-00079, Artificial Intelligence Graduate School Program, Korea University).}
}

\author{\IEEEauthorblockN{Gi-Hwan Shin}
\IEEEauthorblockA{\textit{Dept. Brain and Cognitive Engineering} \\
\textit{Korea University}\\
Seoul, Republic of Korea \\
gh\_shin@korea.ac.kr}
\and
\IEEEauthorblockN{Young-Seok Kweon}
\IEEEauthorblockA{\textit{Dept. Brain and Cognitive Engineering} \\
\textit{Korea University}\\
Seoul, Republic of Korea \\
youngseokkweon@korea.ac.kr}
}

\maketitle
\begin{abstract}
Many studies have discussed the difference in brain activity related to encoding and retrieval of working memory (WM) tasks. However, it remains unclear if there is a change in brain activation associated with re-encoding. The main objective of this study was to compare different brain states (rest, encoding, and re-encoding) during the WM task. We recorded brain activity from thirty-seven participants using an electroencephalogram and calculated power spectral density (PSD) and phase-locking value (PLV) for different frequencies. In addition, the difference in phase-amplitude coupling (PAC) between encoding and re-encoding was investigated. Our results showed that alpha PSD decreased as the learning progressed, and theta PLV, beta PLV, and gamma PLV showed differences between brain regions. Also, there was a statistically significant difference in PAC. These findings suggest the possibility of improving the efficiency of learning during re-encoding by understanding the differences in neural correlation related to learning.
\end{abstract}


\section{Introduction}
Working memory (WM), defined as the ability to maintain and manipulate relevant information for a short period of time, is a fundamental component of cognitive activities such as learning and problem-solving \cite{d2015cognitive}. An important goal of WM cognitive neuroscience is to identify the brain mechanisms underlying the encoding and retrieval processes \cite{small2001circuit}. Neural activities associated with memory formation can be observed using various devices, including functional magnetic resonance imaging for BOLD activity \cite{zhang2017hybrid, zhang2019strength} and electroencephalogram (EEG) for electrophysiology activity \cite{jeong2020brain,lee2018high}. The measurement with a high temporal resolution is required to extract the characteristics of rapidly fluctuating brain activation \cite{kwon2019subject}. Therefore, we recorded the EEG during the WM task to analyze the neural coupling involved in learning.  

In previous studies, various EEG analyses were applied to identify differences according to brain regions and frequencies during WM tasks. First, the power spectral density (PSD) analysis to confirm brain activation showed alpha and beta power decreased while theta and gamma power increased \cite{hanslmayr2012oscillatory}. Second, it was confirmed that there was a significant correlation between frequency bands and brain regions during memory encoding through the functional connectivity indicators representing the quantitative phase relationships of memory processes \cite{chou2015explore}. Finally, in the phase-amplitude coupling (PAC), which is a method for calculating various forms of neural synchronization between oscillations across different frequency bands, both increases and decreases in PAC during memory encoding were found to be relevant \cite{munia2019time,vaz2017dual}. These results showed differences in EEG signals in relation to memory encoding. However, it remains unclear about brain functional differences related to re-encoding. 

In this study, we aimed to compare neural activation of different brain states (rest, encoding, and re-encoding) during working memory. In particular, we focused on identifying differences in the relationship between phase and amplitude interactions during encoding and re-encoding. Participants performed paired associated memory task. We calculated brain activation using PSD, PLV, and PAC in the measured EEG data. Our findings could be a tool to improve memory performance in relation to encoding and re-encoding in WM task.

\begin{figure*}[t!]
\centering
\scriptsize
\includegraphics[width=\textwidth]{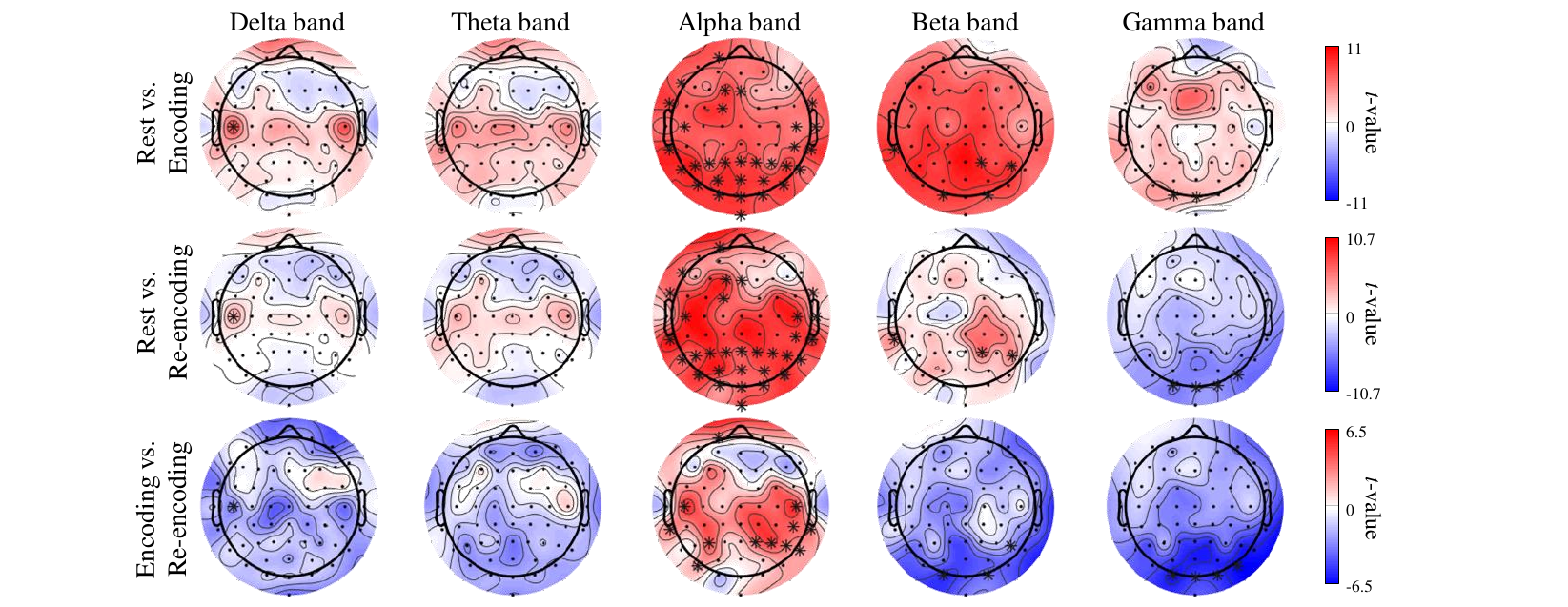}
\caption{Statistical differences of spectral power in each frequency band between different types. The red regions mean the latter is smaller than the former, while the blue regions mean the latter is larger than the former. The black asterisk indicates a significant electrode (\textit{p} $<$ 0.05 with Bonferroni's correction).}
\end{figure*}

\section{Materials and Methods}

\subsection{Participants}
Thirty-seven healthy participants underwent in this study (21 males and 16 females; age 25.22 $\pm$ 2.37 years). This study was approved by the Institutional Review Board at Korea University (KUIRB-2021-0155-03) and written informed consent was obtained from participants before the experiment.

\subsection{Working Memory Task} 
The word-pair memory task consisted of learning and recall sessions was previously covered by Shin \textit{et al.} \cite{shin2020assessment}. All participants were asked to memorize 54 semantically related word pairs \cite{marshall2006boosting}. In the learning session, each pair of words was displayed on a monitor for 4 sec in random order, followed by a 1 sec rest. In the recall session, the characters of word corresponding to the word displayed on the screen were entered on the keyboard within 30 sec and re-encoding was presented for 2 sec after the response. The task was implemented using Psychtoolbox (http://psychtoolbox.org).

\subsection{Data Recording and Preprocessing}
We obtained signals from the scalp and face with 60-channels EEG and 4-channels electrooculogram (EOG) of Ag/AgCI electrode according to the 10-20 international system using BrainAmp (ActiCap, Brain Products, Germany). The reference and ground electrodes were placed on FCz and AFz, respectively. The sampling frequency of all electrodes was 1,000 Hz and the impedance was kept below 20 k$\Omega$.

Measured data were processed in MATLAB 2018b using the EEGLAB toolbox \cite{delorme2004eeglab} and BCILAB toolbox \cite{kothe2013bcilab}. The raw signals were down-sampled to 250 Hz, band-pass filtered between 0.5 to 100 Hz, and re-referenced to the average reference. Then, spatial noise was filtered using surface Laplacian and EOG artifacts were removed from EEG data using independent component analysis. The preprocessed EEG data were segmented into three different types of epochs: 1) 1 sec rest, 2) 4 sec encoding, and 3) 2 sec re-encoding.

\subsection{EEG Data Analysis}
To identify the characteristics of the conditions according to the frequency of interest, we divided it into delta (2-4Hz), theta (4-8 Hz), alpha (8-13 Hz), beta (13-30 Hz), and gamma (30-100 Hz) bands. We also were grouped into five brain regions (frontal, central, temporal, parietal, and occipital regions) to compare differences between brain regions \cite{shin2020assessment}.

\subsubsection{Spectral Power}
To evaluate the distribution of power in signals over frequency bands during different brain states, we calculated PSD using a fast Fourier transform, which converts from the time domain to frequency domain \cite{lee2019possible,suk2014predicting}. For each channel, the spectral power of the five frequency bands was calculated for each participant and averaged over all trials.

\subsubsection{Functional Connectivity}
We calculated a phase-locking value (PLV), which measures the phase stability between pair of channels in the same frequency band for each condition \cite{lee2019connectivity}. For each participant, all trials of 60-channel EEG data were averaged, resulting in a 60 x 60 matrix. Then, the channels were grouped into five brain regions.

\begin{figure*}[t!]
\centering
\scriptsize
\includegraphics[width=\textwidth]{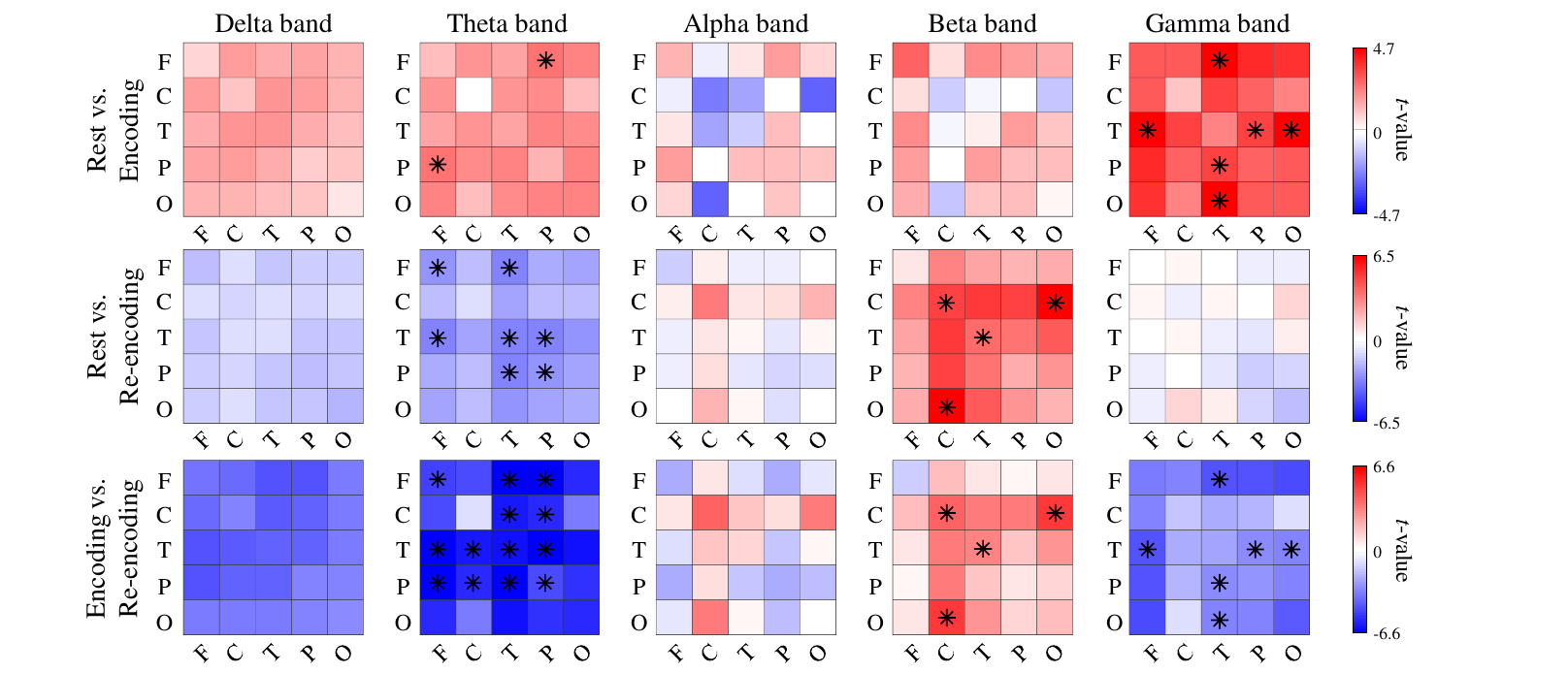}
\caption{Statistical differences between brain regions in each frequency band between different types. The red color means the latter is smaller than the former, while the blue color means the latter is larger than the former. The black asterisk indicates a significant connectivity (\textit{p} $<$ 0.05 with Bonferroni's correction).}
\end{figure*}

\begin{figure*}[t!]
\centering
\scriptsize
\includegraphics[width=\textwidth]{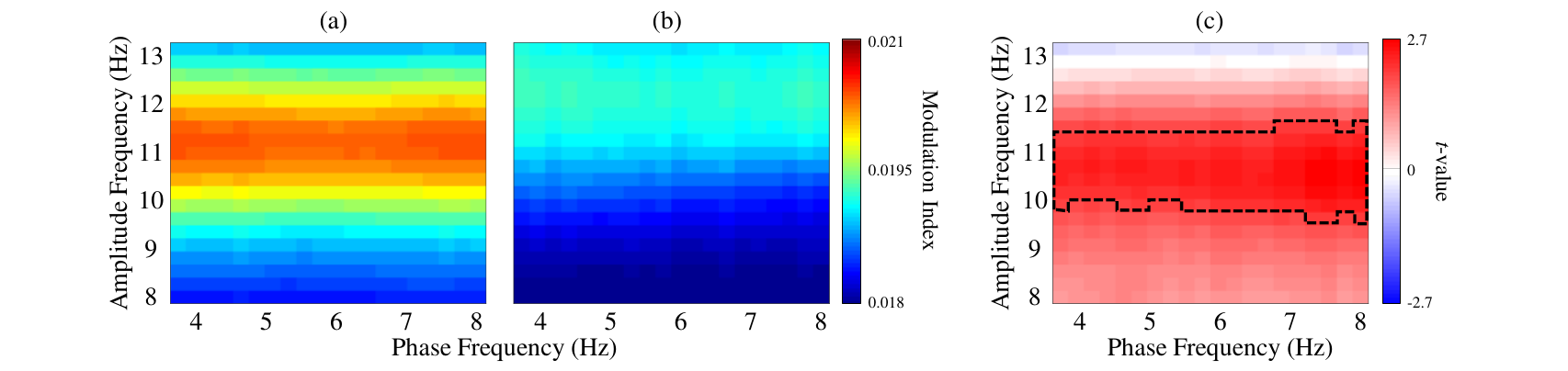}
\caption{Comodulogram of theta-alpha phase-amplitude coupling for (a) encoding, (b) re-encoding, and (c) encoding vs. re-encoding. It represents an averaged modulation index calculated between phase frequencies from 4 to 8 Hz and amplitude frequencies from 8 to 13 Hz. The dashed line indicates a statistically significant difference between the two conditions (\textit{p} $<$ 0.05).}
\end{figure*}

\subsubsection{Phase-Amplitude Coupling}
The PAC analysis was performed based on frequencies and brain regions that showed significant results in the above two methods. We used the modulation index (MI), which measures cross-frequency coupling based on one of the PAC methods, Shannon entropy and Kullback-Leibler divergence \cite{tort2008dynamic, lee2020frontal}. The MI calculation is detailed in tort \textit{et al}. \cite{tort2008dynamic}. The MI is reported to be relatively insensitive to changes of noise level and short epochs \cite{hulsemann2019quantification}. To compare the differences between encoding and re-encoding, we computed the MI for the corresponding frequency pairs. 

\subsection{Statistical Analysis}
To investigate statistical differences among brain states during working memory, a one-way analysis of variance was applied to each of PSD and PLV. For post-hoc analysis, a paired sampled \textit{t}-test with Bonferroni's correction was performed to identify differences between conditions. We also performed a paired sampled \textit{t}-test with Bonferroni's correction to compare MI of PAC in encoding vs. re-encoding. The alpha level for all statistical analyses was set at 0.05.

\section{Results}
\subsection{Difference in Spectral Power by Brain States} 
We examined the difference in spectral power according to frequency bands in different states (Fig. 1). Similar results were derived from all comparisons, and changes in alpha PSD were particularly noticeable. Majority of channels belonging to the parietal and occipital regions between rest vs. encoding and rest vs. re-encoding decreased compared to rest, and re-encoding decreased in the parietal region than encoding. On the other hand, there was no difference in the theta PSD in all comparisons. In summary, the alpha power decreased as encoding and re-encoding progressed. 

\subsection{Difference in Functional Connectivity by Brain States} 
Differences in connectivity were explored between brain regions according to each frequency band (Fig. 2). In rest vs. encoding states, there was a statistical difference in theta PLV of the frontal-parietal region and gamma PLV of temporal and other regions except for the central region. In rest vs. re-encoding states, theta PLV increased in frontal, temporal, and parietal regions, while beta PLV decreased in central, central-occipital, and temporal regions. Finally, in encoding vs. re-encoding, theta PLV was related to the frontal, temporal, and parietal regions, and gamma PLV was related to the temporal and other regions. In addition, beta PLV showed differences in the central, central-occipital, and temporal regions. In all comparisons, delta PLV and alpha PLV had no statistical difference between brain regions.

\subsection{Difference in Phase-Amplitude Coupling of Two Conditions} 
We investigated changes in theta-alpha PAC based on the PSD and PLV results. In particular, we compared the differences between the theta phase in the temporal-parietal regions and the alpha amplitude in the parietal region during encoding and re-encoding. As a result, the highest PAC was found at 4-8 Hz phase and 10-12 Hz amplitude in encoding (Fig. 3a), and the highest PAC was found at 4-8 Hz phase and 11-13 Hz amplitude in re-encoding (Fig. 3b). Overall, the PAC of encoding was higher than the re-encoding. When the statistics between the two conditions, there was a significant difference at the 4-8 Hz phase and 10-12 Hz amplitude (Fig. 3c).

\section{Discussion}
In this study, we compared brain activation of EEG segments of rest, encoding, and re-encoding during WM task. As a result, as encoding progressed, alpha PSD significantly decreased in temporal and parietal regions, and changes were observed between brain regions in theta PLV, beta PLV, and gamma PLV. Also, statistical differences between theta-alpha PAC of learning and re-learning were identified.

Previous studies have investigated the role of alpha oscillations and found that alpha PSD decreased during WM, especially in the posterior region \cite{bashivan2014spectrotemporal}. Indeed, asynchronous neural activity reflected by decreased alpha power was found to be positively associated with the memory process \cite{hanslmayr2012oscillatory}. Thus, our observations were consistent with the previous literature. In addition, we observed differences in connectivity between brain regions. Functional interactions between neurons in the brain can explain mechanisms of information processing \cite{funahashi2006prefrontal}. For example, it has been reported that large-scale theta and gamma synchronization during WM task in encoding are associated with co-activation of cortical networks and that the connectivity of memory formation processes in the frontal, temporal, and parietal regions is prominent. \cite{xie2021theta}. Therefore, these results confirmed that there was a difference between memory processes during the WM task.

Finally, we found that the PAC of re-encoding was significantly lower than the PAC of encoding. Most interpretations of previous studies have suggested that PAC increases are functionally important \cite{tort2008dynamic,lee2020frontal}, but recent studies have also demonstrated a decreased role of PAC for memory formation in the human hippocampus \cite{leszczynski2015rhythmic,hanslmayr2016oscillations}. Specifically, Hanslmayr \textit{et al}. \cite{hanslmayr2016oscillations} suggested the possibility that the neuronal over-coupling by PAC may play a role in inhibiting rather than enhancing effective neural processing. Thus, these results showed that different brain mechanisms appeared between encoding and re-encoding.

The limitation of this paper is the experimental paradigm consists of only one re-encoding. It is necessary to understand the difference more accurately through experiments on re-encoding according to the number of trials and time. In future works, additional experiments will be conducted to confirm the possibility of improving memory performance by applying encoding and re-encoding feedback in real-time based on differences in brain activation identified in WM task.

\section{Conclusion}

In the current study, we compared the brain activation patterns of three EEG types: rest, encoding, and re-encoding. As encoding progressed, statistical differences in PSD and PLV were identified according to the brain regions and frequencies. Also, encoding vs. re-encoding, the difference in PAC could be confirmed. These findings could help improve cognitive function through applications such as neurofeedback training by extracting features from re-encoding in real-time. 


\bibliographystyle{IEEE}
\bibliography{reference}

\end{document}